\documentclass[iicol, pdflatex,sn-basic]{sn-jnl}

\usepackage{soul}
\usepackage{amsmath}
\usepackage{subfig}
\usepackage{float}
\usepackage{array}
\usepackage{multirow}
\usepackage{comment}
\usepackage{adjustbox}
\usepackage{graphicx}

\jyear{2022}%

\theoremstyle{thmstyleone}%
\theoremstyle{thmstyletwo}%

\theoremstyle{thmstylethree}%

\newcommand{\ijdl}[1]{\textcolor{black}{#1}}

\begin{document}

\title[Retrievability in an IR System]{Retrievability in an Integrated Retrieval System: An Extended Study}



\author*[1]{\fnm{Dwaipayan} \sur{Roy}}\email{Dwaipayan.Roy@iiserkol.ac.in}

\author[2]{\fnm{Zeljko} \sur{Carevic}}\email{Zeljko.Carevic@gesis.org}

\author*[2]{\fnm{Philipp} \sur{Mayr}}\email{Philipp.Mayr@gesis.org}

\affil*[1]{\orgname{Indian Institute of Science Education and Research}, \orgaddress{ \city{Kolkata}, \postcode{741246},  \country{India}}}

\affil*[2]{\orgname{GESIS -- Leibniz Institute for the Social Sciences}, \orgaddress{\street{Unter Sachsenhausen 6-8}, \city{Cologne}, \postcode{50667}, \country{Germany}}}


\abstract{
\ijdl{Retrievability measures the influence a retrieval system has on the access to information in a given collection of items.
This measure can help in making an evaluation of the search system based on which insights can be drawn.
In this paper, we investigate the retrievability in an integrated search system consisting of items from various categories,
particularly focussing on datasets, publications \ijdl{and variables} in a real-life Digital Library (DL). 
The traditional metrics, that is, the Lorenz curve and Gini coefficient, are employed to visualize the diversity in retrievability scores of the \ijdl{three} retrievable document types (specifically datasets, publications, and variables).
Our results show a significant popularity bias with certain items being retrieved more often than others. Particularly, it has been shown that certain datasets are more likely to be retrieved than other datasets in the same category. In contrast, the retrievability scores of items from the variable or publication category are more evenly distributed. We have observed that the distribution of document retrievability is more diverse for  datasets as compared to publications and variables.}
}

\keywords{Retrievability, Dataset Retrieval, Interactive IR, Diversity}



\maketitle

\section{Introduction}

In the present era of information, we are generating a colossal amount of data that needs to be handled and processed efficiently for quick look-ups.
The expeditious advancement in technologies has made data generation even more complex with a diversified form of information coming from divergent sources.
This necessitates the need to have a federated or integrated system~\citep{adali:icde1995, federated-arguello} that searches and assimilates results from assorted sources.
Textual data still remains the predominant type among them and significant research has been conducted in the domain of textual document retrieval. 
Among the rest, recent research on dataset retrieval~\citep{dataset-ret} has become increasingly important in the (interactive) information retrieval and digital library communities. One of the reasons is undoubtedly the enormous number of research datasets available. However, the underlying characteristics of dataset retrieval also contribute to the attention in this area. One often-mentioned characteristic is the increased complexity of datasets over traditional document retrieval. While the latter is well-known and adequately studied, datasets often include more extensive material and structures that are relevant for retrieval. This may involve the raw data, descriptions of how the data was collected, taxonomic information, questionnaires, codebooks, etc. 
Recently, numerous studies have been conducted to further identify the characteristics of dataset retrieval. These studies include the observation of data retrieval practices \citep{kramer2021data}, interviews and online questionnaires \citep{kern2015there,friedrich2020looking} and transaction log analysis \citep{kacprzak2017query,carevic2020characteristics}.

In this paper, we follow a system-oriented approach for studying dataset retrieval. By employing the measure of \em retrievability \em \citep{ret-cikm08}, we aim to gain insights into the particularities of dataset retrieval in comparison to traditional document retrieval. The measure of retrievability was initially developed to quantify the influence that a retrieval system has on access to information. In a simplified way, retrievability represents the ease with which a document can be retrieved given a particular IR system \citep{ret-cikm08}. The measure of retrievability can be utilised for several use cases. 

\ijdl{As an extension of our prior work~\citep{royjcdl22}, } we investigate the retrievability of various types of documents in an integrated digital library \textit{GESIS Search} (see Section \ref{sec:ret-hetero}), focusing on various types of data, particularly datasets, publications \ijdl{and variables}. The assumption followed here is that in an ideal ranking system\footnote{In this paper, by \emph{ranking system} or, \emph{IR system}, we refer to \emph{a system} containing a corpus together with the retrieval model to be used to search on that corpus.}, the retrievability of each indexed item (dataset or other publication) is equally distributed. Likewise, a discrepancy to this assumption may reveal an inequality between the items in a collection caused by the system. By employing a measure of retrievability, we expect to gain further insight into the characteristics of dataset retrieval compared to traditional document retrieval.


\subsection*{Research questions}
\ijdl{We verify the research questions put forward and discussed by~\cite{royjcdl22} in the updated system with a variety of item types tested with more queries (see Section \ref{sec:exp}).
Similar to the previous work, we substantiate the following research questions on the integrated search system \textit{GESIS Search} focusing on an additional type of item: \emph{Variables} together with \emph{Publication} and \emph{Dataset}}:
\begin{itemize}
    \item \textbf{RQ1:} In the integrated search system with various types of items, can we observe any prior bias of accessibility of documents from a particular type?
    \item \textbf{RQ2:} Can we formalize this type-accessibility bias utilizing the concept of document retrievability?
    \item \textbf{RQ3:} How diverse are the retrievability score distributions in the different categories of documents in our integrated search system?
\end{itemize}

\ijdl{Our previous study \citep{royjcdl22} was designed to take all queries in the query log into account. This had the benefit of being as close to the real search behaviour as possible. At the same time this design choice introduced a popularity bias caused by reoccurring queries that positively influence the retrievability score of documents in the corresponding result set.
Additionally, the popularity bias of queries has been ignored in this work.
Thus, contrasting with the previously reported results, we address the following research question: 
\begin{itemize}
    \item \ijdl{\textbf{RQ4:} In a real-life search system, does popularity bias of queries influence the inequality in any way?} 
\end{itemize}
}

In sum, our contributions are as follows: 1) we utilize the retrievability measure to better understand the diversity of accessing datasets in comparison to publications with real-life queries from a search log; 2) building on retrievability, we propose \ijdl{to employ} the measurement of \em usefulness\em, which represents implicit relevance signals observed for datasets and publications. Our understanding of bias follows the argumentation provided in \cite{wilkie2017algorithmic} where bias denotes the inequality between documents in terms of their retrievability within the collection. Bias can be observed when a document is overly or unduly favoured due to some document features (e.g. length, term distribution, etc.) \citep{wilkie2014best}.

The rest of the paper is organized as follows.
We first present background and related work in Section~\ref{sec:rel-work} together with formally introducing the concept of retrievability. 
The integrated search system \textit{GESIS Search} along with the motivation of our retrievability study is presented in Section~\ref{sec:ret-hetero}.
Section~\ref{sec:exp} discusses the empirical results and analysis of the outcome of the experimentation before introducing the novel concept of usefulness in Section~\ref{sec:usefulness} along with the experimental study of usefulness.
We conclude the paper in Section~\ref{sec:conclusion} highlighting the contributions and findings of the paper with directions to extend the work.

\section{Background and related work}\label{sec:rel-work}
\ijdl{
Considering a collection of items, the retrievability of items can be defined as how accessible or findable the items are by some searching techniques. 
In context of document retrieval, the concept was developed and proposed in~\cite{ret-cikm08}.
Informally, the retrievability of a document in a collection indicates the expectation of selection of the document by some retrieval model within a rank cutoff.
Mathematically, the retrievability of any document $d$ in a collection $C$ is defined as:
\begin{equation}\label{eq:ret1}
    r(d) = \sum_{q\in Q} w_q \cdot f(rank(d,q,M),c)
\end{equation}
where, 
\begin{itemize}
    \item $Q$ - the set of all queries which are answerable by the collection;
    \item $w_q$ - weight of the query $q$;
    \item $rank(d,q,M)$ - rank of the document $d$ when retrieval is performed with query $q$ using retrieval model $M$;
    \item $c$ - the rank cutoff.
\end{itemize}
The function $f(rank(d,q,M),c)$ is an \emph{indicator function} that returns either $1$ or $0$ depending on whether the rank ($rank(d,q,M)$) of document $d$ is within the rank cutoff $c$ or not.
The indicator function can be mathematically defined as the following:
\begin{equation}\label{eq:fk}
  f(rank(d,q,M),c)=\begin{cases}
    1, & \text{if $rank(d,q,M) \leq c$}.\\
    0, & \text{otherwise}.
  \end{cases}
\end{equation}
\newline
In Equation~\ref{eq:ret1}, the retrievability of a document is computed based on retrieval performed with all sets of queries $\mathsf{Q}$ addressable by the document collection. 
Considering a sizeable collection of documents, there can be infinitely many distinct queries that can be answered by various documents in the collection.
One of the practical approaches to get this set of all queries $\mathsf{Q}$ is to use a query log; however acquiring such a log is not always feasible. 
In the absence, a query-based sampling method~\citep{callan2001query} can be applied to randomly populate $\mathsf{Q}$.
In~\cite{ret-cikm08}, the authors considered generating queries with unigrams and bigrams based on the collection frequency of them above a threshold in the collection.
This approach may result in an enormous number of queries if a large collection of documents is considered. 
To keep the experimental setup tractable, one approach here is to truncate the list again based on a certain threshold value (e.g. 2 million as selected by~\citeauthor{ret-cikm08}). 
Hence, the construction of $\mathsf{Q}$ based on either query log or random sampling of terms from the collection are some practical approximations that we can adapt in order to realize the concept of retrievability of documents in a collection.
\newline
The query weight $w_q$ in Equation~\ref{eq:ret1} may be used for incorporating a bias (such as popularity, importance, etc.) in the retrievability computation.
Ignoring these biases, this weight is considered as uniform for all queries in earlier works~\citep{ret-cikm08, bashir2009analyzing,bashir2009CIKM}. 
The approximated retrievability score ($\hat{r}(d)$) of document $d$ will then be a discrete value $x$ indicating the number of queries for which $d$ is retrieved within top rank $c$.
Certainly, this is a simplifying assumption and the queries submitted to a search system in practice vary vastly both in terms of \emph{popularity} as well as \emph{difficulty}~\citep{query-dif2006carmel}.
\newline
The second factor of the per-query component in Equation~\ref{eq:ret1} is a boolean function that depends solely on the rank at which document $d$ is retrieved.
Increasing the value of the rank cut-off ($c$) broadens the domain of documents retrieved which will positively influence the retrievability scores of more documents.
Note that being selected by a retrieval model for some queries does not ensure the relevance of the document which can only be assessed by human judgements. 
}

Retrievability as a measure was proposed in \cite{ret-cikm08} where the authors experiment on two TREC collections with queries generated using a query-based sampling technique~\citep{callan2001query}. 
Since then, Retrievability has been primarily used to detect bias in ranking systems. For instance, \cite{samar2018quantifying} employ retrievability to research the effect of bias across time for different document versions (treated as independent documents) in a web archive. Their results show a ranking bias for different versions of the same document.  Furthermore, the study confirms a relationship between retrievability and findability measured by Mean Reciprocal Rank (MRR). They follow the assumption that the lower a document's retrievability score the more difficult it is to find the document.
Another application of the retrievability measure can be found in patent or legal document retrieval which provides a unique use case due to its recall-oriented application. 
In both studies \citep{bashir2009analyzing,bashir2009CIKM}, the authors look at document retrievability measurements and argue that a single retrievability measure has several limitations in terms of interpretability. In \cite{bashir2009analyzing} they try to improve accessibility measurement by considering sets of relevant and irrelevant queries for each document. In this way, they try to simulate recall-oriented users. In addition, they plot different retrievability curves to better spot the gaps between an optimal retrievable system and the tested system. The other work \citep{bashir2009CIKM} analyze the bias impact of different retrieval models and query expansion strategies. Their experiments show that clustering-based document selection for pseudo-relevance feedback is an effective approach for increasing the findability of individual documents and decreasing the bias of a retrieval system.
Further researches on patent retrieval reported in \cite{bashir2009pair} and \cite{bache2010patent} identify content-based features that can be used to classify a set of documents based on their retrievability. Experiments on various patent collections show that these features can achieve more than 80\% classiﬁcation accuracy.

 A study on the query list generation phase for determining the measure of retrievability is presented in \cite{bashir2011relationship}. The study addresses two central problems when determining retrievability: 1) query selection and 2) query characteristics identification. It is argued that the query selection phase is usually performed individually without well-accepted criteria for query generation. Hence their goal is to evaluate how far the selection of query subsets provides an accurate approximation of retrieval bias. The second shortcoming is addressed by determining retrievability bias considering different query characteristics. In their experiments, they recognise that query characteristics influence the increase or decrease of retrievability scores. 
 A topic-centric query generation technique, tested on the Associated Press (AP) document collection, is proposed in~\cite{wilkie2016topical}.
A significant correlation is reported between the traditional estimate of Gini and the estimate produced by this method of topic-centric query.
 As recognised in \cite{bashir2011relationship}, the majority of retrievability experiments employ simulated queries to determine retrievability. To study the ability of the retrieval measure in detecting a potential retrievability bias using real queries issued by users, \cite{TraubSOHVH16} conducted an experiment on a newspaper corpus. Their study confirms the ability to expose retrievability bias within a more realistic setting using real-world queries. A comparison of simulated and real queries with regard to retrievability scores further shows considerable differences which indicate a need for improved construction of simulated queries.
 To see if there is any correlation between the retrievability bias and performance measurement, in another study, \cite{wilkie2014retrievability} examine the relationship between retrieval bias and ten retrieval performance measures. Experimentation of TREC ad-hoc data demonstrates that the retrievability bias hypothesis tends to hold for most of the performance measurements.

Retrievability of documents indicates the chance of selection by a retrieval model for various queries submitted. 
However, the selection of a document does not mean that the document is indeed \textit{useful} in addressing the information need generating the query.
This can only be realised by using document consumption signals (e.g. in the form of relevance judgements).
This concept was first introduced in \cite{cole2009usefulness} as a criterion to determine how well a system is able to solve a user's information need.
In their work, \citeauthor{cole2009usefulness} denoted this notion as \textit{usefulness}.
 In \cite{hienert2016usefulness}, it has been operationalised within a log-based evaluation approach to determine the usefulness of a  search term suggestion service. 
 The usefulness has been further operationalised in \cite{carevic2018contextualised} to determine the effects of contextualised stratagem browsing on the success of a search session. 




Recently, a considerable amount of research has been carried out concerning the characteristics of dataset retrieval. A comprehensive literature review on dataset retrieval is provided in \cite{gregory2019searching} focusing on dataset retrieval practices in different disciplines.  
Research in this area covers, for instance, the analysis of information-seeking behaviour during dataset retrieval through observations \cite{kramer2021data}, questionnaires and interviews \citep{kern2015there,friedrich2020looking}, and transaction-log studies \citep{kacprzak2017query,carevic2020characteristics}. 
In~\cite{kern2015there}, the authors investigated the requirements that users have for a dataset retrieval system. Their findings on dataset retrieval practices suggest that users invest greater effort during relevance assessment of a dataset. They conclude that the selection of a dataset is a much more important decision compared to the selection of a piece of literature. This results in high demands on  metadata quality  during the dataset retrieval. The complexity of assessing the relevance of a dataset is also highlighted in \cite{kramer2021data}.  
Besides topical relevance, access to metadata as well as documentation about the dataset plays a crucial role. A query log analysis from four open data portals is presented in \cite{kacprzak2017query}. Their study indicates differences between queries issued towards a dataset retrieval system and queries in web search. 
In a subsequent study \citep{kacprzak2018characterising}, the extracted queries are further compared to queries generated from a crowdsourcing task. The intuition and focus of this work is to determine whether queries issued towards a data portal differ from those collected in a less constrained environment (crowdsourcing).

\section{Retrievability in an integrated retrieval system}\label{sec:ret-hetero} 

\begin{figure*}
    \centering
    \includegraphics[scale=0.43]{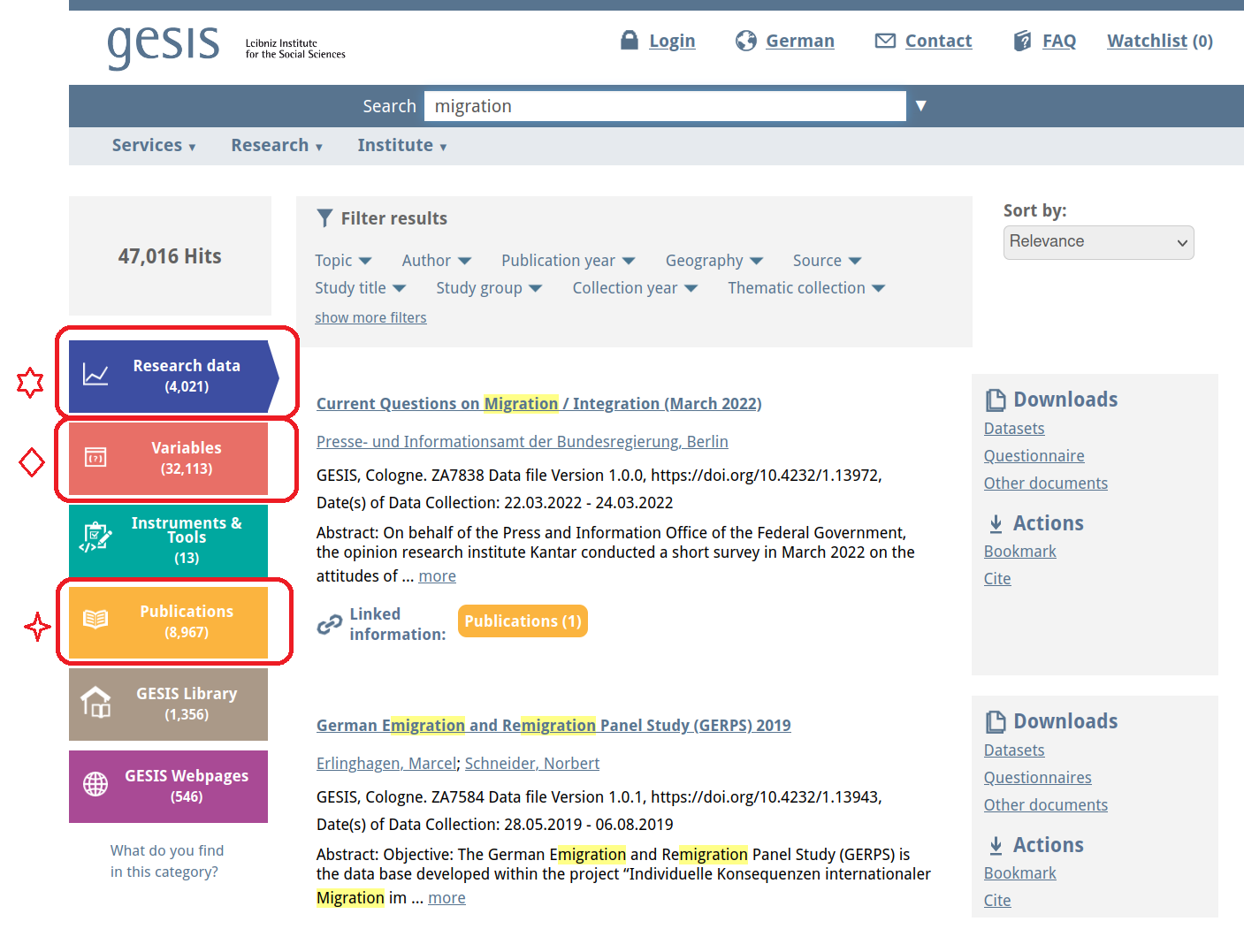}
    \caption{Screenshot of GESIS Search showing result sets for research data, publications, and variables.}
    \label{fig:gesis-search-sc}
\end{figure*}

\begin{figure*}
    \centering
    \includegraphics[scale=0.3]{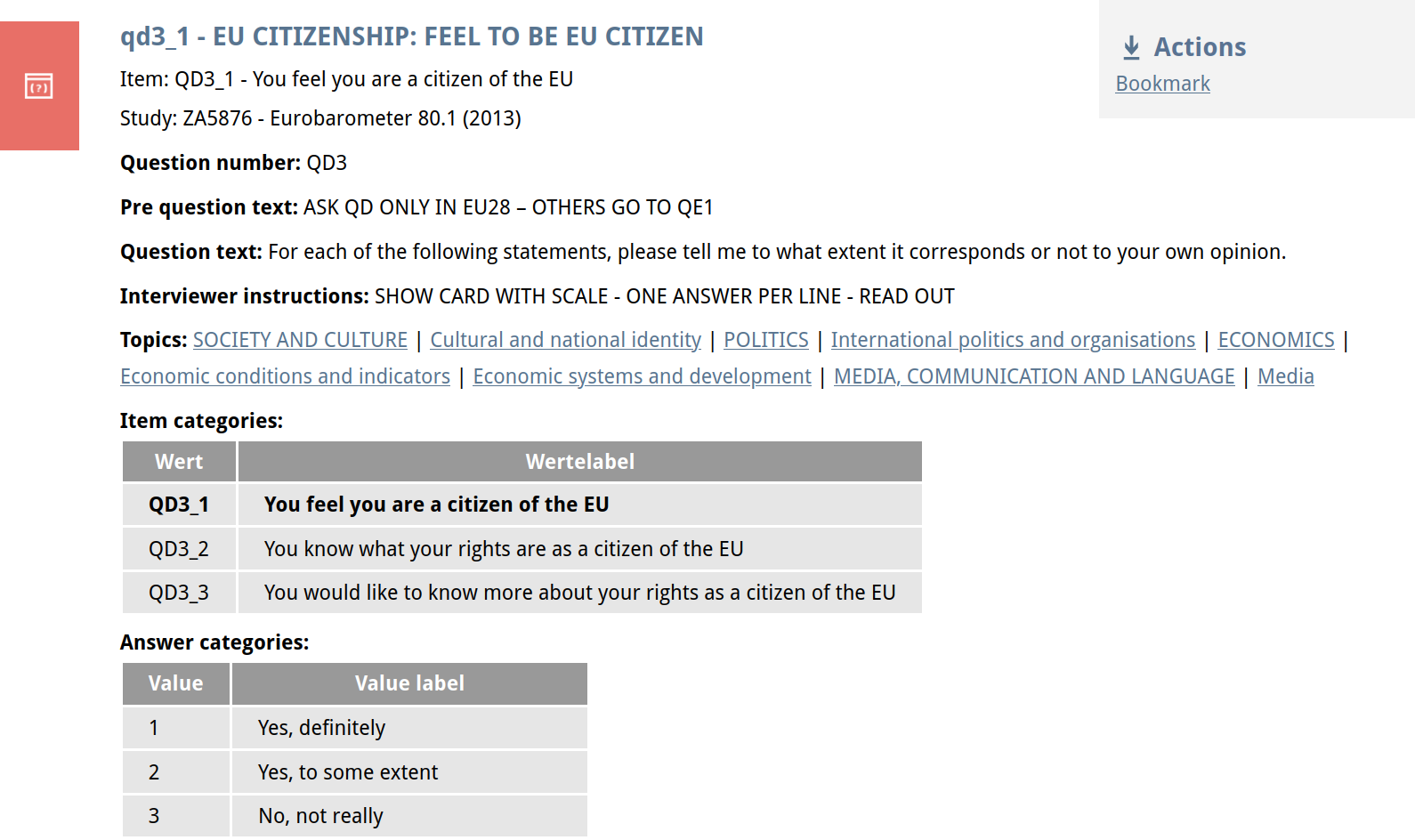}
    \caption{Screenshot of the variable description of variable \texttt{QD3\_1}  in the GESIS Search.}
    \label{fig:gesis-search-variable}
\end{figure*}

We define an \emph{integrated search system} as a system that searches multiple sources of different types and integrates the output in a unified framework\footnote{This is similar to the concepts of aggregated search~\citep{agreegated-lalmas} or federated search~\citep{federated-arguello}.}.
The retrieval in such a system requires sophisticated decision-making considering the various modalities in documents in the collection of data.

Following Equation~\ref{eq:ret1}, the retrievability score of documents is dependent on the other documents in the collection\footnote{Here, we are considering the employed retrieval function as constant.}:
considering a rank-cut $c$, the rank of a document under consideration can be greater than $c$ ($>c$) due to the documents, taking the top $c$ positions, being more relevant or duplicate~\citep{Nikkhoo2011TheIO}.
Another factor that can influence the retrievability score of a document is its popularity; a popular document will be retrieved multiple times by users over time.
In case of an integrated search engine, where the documents belong to various categories, some particular types could be having higher chances than others in terms of being retrieved. 
In general, there can be some disparity in the number of documents of various categories being retrieved which can be a result of popularity bias in the collection. 
This type of popularity bias can impede the satisfaction of the information need of a user, and in turn, can affect the performance of the system.
The satisfaction of a user can only be realised via a direct feedback from them. 
In absence of such explicit information, it is strenuous, if at all possible, to understand whether information need is fulfilled or not.
%
In this article, we are going to present an extended study of the diversity in retrievability scores for different categories of documents in the integrated search system \textit{GESIS Search\footnote{\url{https://search.gesis.org}} } \citep{hienert2019}.


\section{Experimental Study}\label{sec:exp}

As presented in Section~\ref{sec:ret-hetero}, we use the integrated search system with various categories of documents in this work.
In this section, we start by describing the data that we have used in the work along with different statistics of the data; this will be followed by the experimental evaluation of the study.

\subsection{Datasets}\label{subsec:datasets}

\ijdl{
We conduct our experimentation on the integrated search system \textit{GESIS Search} containing a total of 860K indexed records (as of November 2022) in different categories such as Research Data, Publications, Variables, Instruments etc. 
Social science publications that are indexed in \textit{GESIS Search} use and reference survey datasets, containing hundreds or thousands of questions. These questions are using so-called survey variables (variables in the following). From an information retrieval perspective, variables in GESIS Search are information objects like datasets with specific metadata elements such as question text, answer categories and frequency tables. }

\ijdl{
A screenshot showing the interface of GESIS search is presented in Figure \ref{fig:gesis-search-sc}. See an example of a variable description in Figure \ref{fig:gesis-search-variable} and the according link to the variable record QD3\_1\footnote{\url{https://search.gesis.org/variables/exploredata-ZA5876\_Varqd3\_1}} in GESIS Search\footnote{Further explanation and examples of Social Science variables and its utilisation for information retrieval can be found in \citep{tsereteli2022}. }. 
The indexed records in GESIS Search are divided into six categories based on their types, covering more than 122K \emph{publications}, 64K \emph{research data} (also referred to as \emph{datasets}), and more than 520K Variables. 
Given a query, the system returns six search result pages (SERP) corresponding to each of the categories (see Figure \ref{fig:gesis-search-sc}). 
The segregation of the SERP enables us to study the retrievability of the different types.
In this study, we specifically focus on the three categories having the largest number of entries, that is, \emph{dataset}, \emph{publications}, and \emph{variables}. 
}

\ijdl{
In the integrated search system, the interaction of the users with the system is logged and stored in a database. 
A total of more than 40 different interaction types are stored covering, for instance, searches (queries), record views and export interactions etc.~\citep{hienert2019}. 
The export of a record belongs to an umbrella of categories including various interactions such as bookmarking, downloading or citing. These interactions are specifically useful for the application of implicit relevance feedback as they indicate a relevance of a record that goes beyond a simple record view. 
The interaction log of the search system provides the basis for our analysis in Section~\ref{subsec:ana_ret} (and later in Section~\ref{subsec:ana_usefulness}). 
These real-user queries form the basis of determining the retrievability of documents.
This ensures realistic queries in $\mathsf{Q}$ of Equation~\ref{eq:ret1} as opposed to the simulated queries used in~\cite{ret-cikm08} or~\cite{TraubSOHVH16}.
The data used in this study is an extended version of our previous work~\citep{royjcdl22}; in this log, all the interactions of real users with the search system were recorded for a period of more than five years, specifically between July 2017 and July 2022. 
The log records more than 2.3 million queries submitted to the integrated search system.
Detailed statistics regarding the extracted interactions utilized in our study can be found in Table~\ref{tab:ext-int}.
Together with the previous observations for record type Publication and Dataset, we report the results on another category, the Variables.
}

\ijdl{
Repeated queries can influence the retrievability score of a document. Formally, the set of all queries $Q$ in Equation~\ref{eq:ret1} may contain the same queries more than once.
For synthetically generated queries (used by \cite{ret-cikm08} and~\cite{bashir2009CIKM}), this can be avoided by keeping track of the already generated queries.
However, the query log of a real-life search system records all such instances where the same queries are given multiple times by the users. 
This factor additionally introduces popularity bias into the reproducibility of documents in the form of query popularity.
The results and observations reported in our earlier study~\citep{royjcdl22} were based on this type of interaction log.
In order to exclusively understand the reproducibility without the query popularity factor, we have only considered unique queries in this work.
}

\begin{table*}[h]
    \centering
\begin{tabular}{ lcccc } 
 \hline
\multirow{2}{*}{\textbf{Record type}} &\multirow{2}{*}{\textbf{Size}}& \textbf{\#queries}  &\textbf{avg. query} &\multirow{2}{*}{\textbf{\#exports}} \\ 
 && (unique) &\textbf{length} & \\ 
 \hline
 \multirow{2}{*}{\textbf{Publication}} & \multirow{2}{*}{113K} & 1,028,485 & \multirow{2}{*}{2.6} & \multirow{2}{*}{63,577} \\
  &  &  (345,144) &  &  \\
 \multirow{2}{*}{\textbf{Dataset}}& \multirow{2}{*}{64K} & 1,208,108 &\multirow{2}{*}{2.3} &\multirow{2}{*}{142,184} \\
  &  & (268,208) & & \\
\multirow{2}{*}{\ijdl{\textbf{Variables}}}& \multirow{2}{*}{523K} & 79,221  &\multirow{2}{*}{2.1} &\multirow{2}{*}{18,832} \\
 &  &  (23,909) & & \\ \hline
\end{tabular}
    \caption{Statistics of the extracted information belonging to the three selected record types.}
    \label{tab:ext-int}
\end{table*}

\begin{table*}[]
\resizebox{0.99\textwidth}{!}{
\begin{tabular}{rrrrrlrrrrlrrrr}
\hline
\multicolumn{1}{l}{\textbf{Rank}}           & \multicolumn{4}{c}{\textbf{Publication}}                                                                                & \multicolumn{1}{l}{} & \multicolumn{4}{c}{\textbf{Research data}}                                                                              & \multicolumn{1}{l}{} & \multicolumn{4}{c}{\textbf{Variables}}                                                                                  \\ \cline{2-5} \cline{7-10} \cline{12-15} 
\multicolumn{1}{l}{\textbf{cutoff}} & \multicolumn{1}{l}{$\mu$} & \multicolumn{1}{l}{g-$\mu$} & \multicolumn{1}{l}{$\sigma^2$} & \multicolumn{1}{l}{$\sigma$} &                       & \multicolumn{1}{l}{$\mu$} & \multicolumn{1}{l}{g-$\mu$} & \multicolumn{1}{l}{$\sigma^2$} & \multicolumn{1}{l}{$\sigma$} &                       & \multicolumn{1}{l}{$\mu$} & \multicolumn{1}{l}{g-$\mu$} & \multicolumn{1}{l}{$\sigma^2$} & \multicolumn{1}{l}{$\sigma$} \\ \hline
10                              & 27.46                    & 7.20                        & 6554.97                      & 80.96                             &                       & 28.16                    & 6.45                       & 12582.64                     & 112.17                            &                       & 2.52                     & 1.77                       & 12.57                        & 3.55                              \\
20                              & 37.56                    & 10.49                      & 9983.99                      & 99.92                             &                       & 39.28                    & 9.11                       & 20022.23                     & 141.50                             &                       & 2.77                     & 1.91                       & 15.05                        & 3.88                              \\
30                              & 46.13                    & 13.65                      & 12666.31                     & 112.54                            &                       & 48.49                    & 11.63                      & 27404.71                     & 165.54                            &                       & 2.97                     & 2.03                       & 16.98                        & 4.12                              \\
40                              & 53.34                    & 16.88                      & 14975.97                     & 122.38                            &                       & 56.3                     & 14.17                      & 33835.99                     & 183.95                            &                       & 3.13                     & 2.12                       & 18.47                        & 4.30                               \\
50                              & 59.66                    & 20.15                      & 16821.35                     & 129.70                             &                       & 63.52                    & 16.97                      & 40087.10                      & 200.22                            &                       & 3.25                     & 2.20                        & 19.52                        & 4.42                              \\
100                             & 66.80                     & 26.09                      & 17923.59                     & 133.88                            &                       & 90.81                    & 32.88                      & 63517.06                     & 252.03                            &                       & 3.67                     & 2.48                       & 22.68                        & 4.76                             \\
\hline
\end{tabular}
}
\caption{The mean (both arithmetic and geometric), variance and standard deviation of the retrievability values when the rank-cutoff is varied.}\label{tab:ret-values-stats}
\end{table*}


\subsection{Measuring retrievability in a collection}\label{subsec:measure-ret}
%
\ijdl{One way of quantifying the information coverage of a collection is by the count of queries that can be addressed (or answered) by the items in the collection.
From the traditional point of view of a web search, the most sought-after way of composing the queries is using free text where vocabulary terms are used to represent an information need.
In a moderate-sized document collection, an intractable number of queries formed using a free-text format are possible. 
Also due to the significant number of documents that can match a free text query, a boolean matching algorithm is not sufficient; this leads to the development of ranked retrieval that returns an ordered list of items sorted based on their relevance.
}

Considering a traditional document collection $\mathsf{C}$, all the documents are not equally important to a query, hence paving the need to have a ranked retrieval.
Now given a set of all possible queries $\mathsf{Q}$, some documents in $\mathsf{C}$ will be relevant to more queries (depending on the topical coverage of the document) than others which can be measured by the concept of retrievability (see Section~\ref{sec:rel-work}).
Formally with the notion of retrievability, some documents will be having higher $r(d)$ in a collection,
resulting in an unequal distribution of retrievability scores.

Similar types of inequalities are observed in economics and social sciences, and they are traditionally measured using the Gini coefficient or Lorenz curve~\citep{lorenz-gini} which measures the statistical dispersion in a distribution\footnote{Lorenz curve and Gini coefficient are popular in economics to measure of wealth disparity in a community/country.}.

Mathematically, the Gini coefficient (G) of a certain value $v$ in a population $\mathcal{P}$ can be defined as:
\begin{equation}\label{eq:gc}
    G = \frac{\sum_{i=1}^{N}{(2*i-N-1) * v(i)}}{N \sum_{j=1}^{N}{v(j)}}
\end{equation}
where $N$ is the size of the population and $v(i)$ specifies the value of $i^{th}$ item in $\mathcal{P}$.
The Gini coefficient in the population will be between $0$ and $1$ and is proportional to the inequality inherent in the population: higher value of $G$ indicates greater disparity and vice versa.
In other words, a value of $G$ equal to $0$ in Equation~\ref{eq:gc} indicates that all the items in the population are equally probable to be selected whereas higher values of $G$ specify a bias implying that only certain items will be selected.


\begin{figure*}[t]
  \centering
  \subfloat[Changes in mean of r(d)] {\label{subfig:mean}\includegraphics[width=0.248\textwidth]{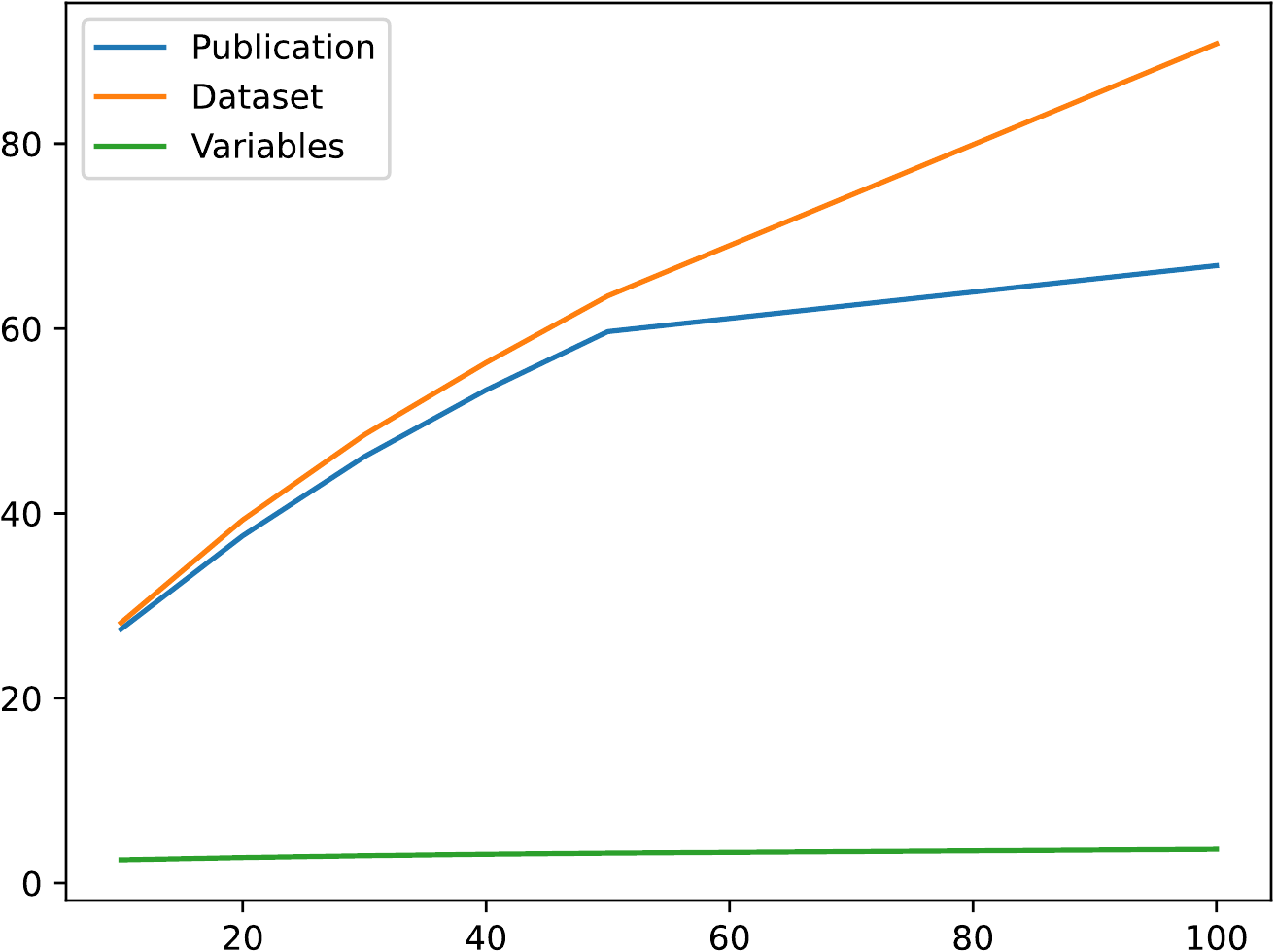}}
  \subfloat[Changes in geometric-mean of r(d)] {\label{subfig:gmean}\includegraphics[width=0.248\textwidth]{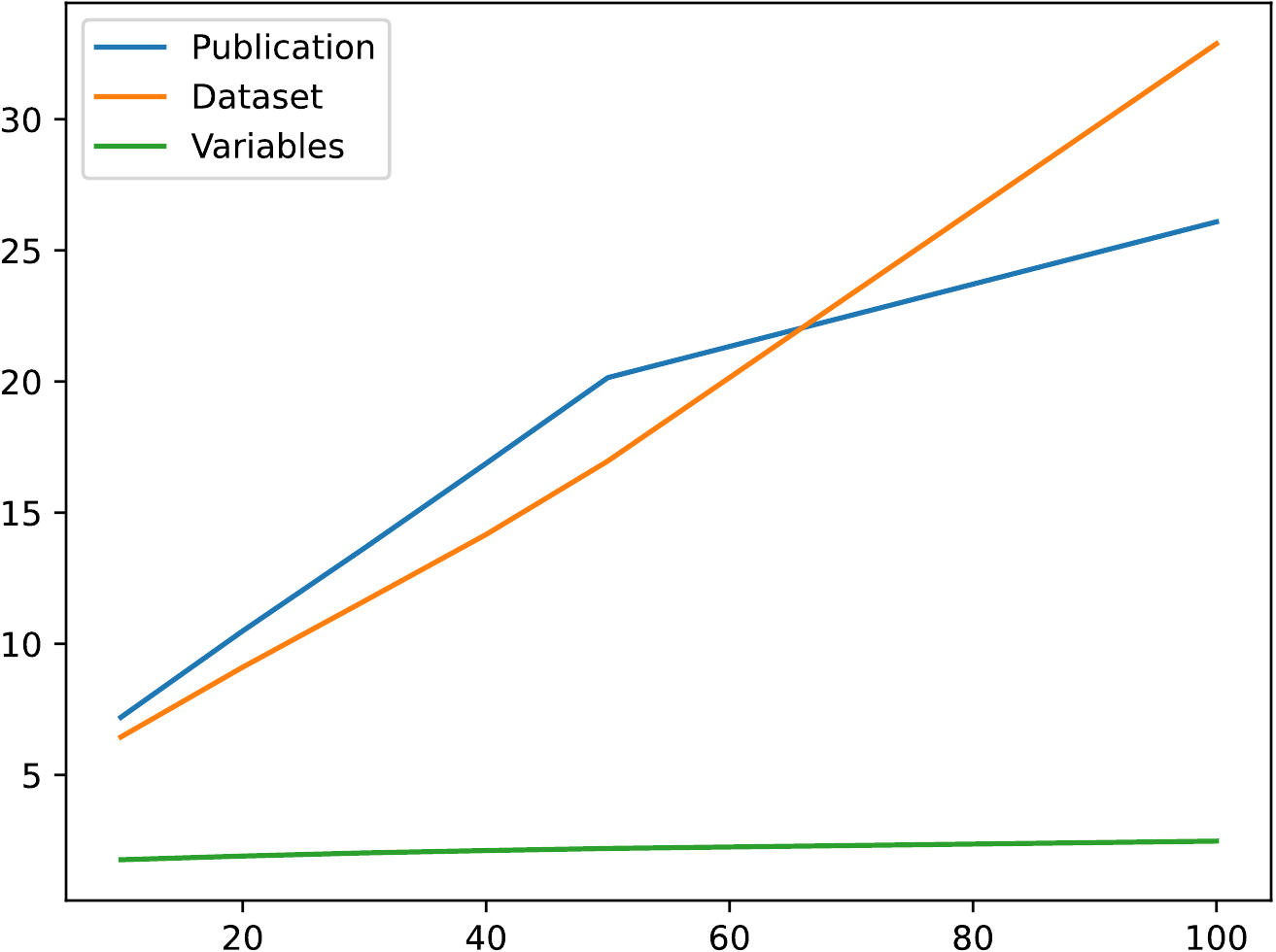}}
  \subfloat[Changes in variance of r(d)] 
{\label{subfig:var}\includegraphics[width=0.26\textwidth]{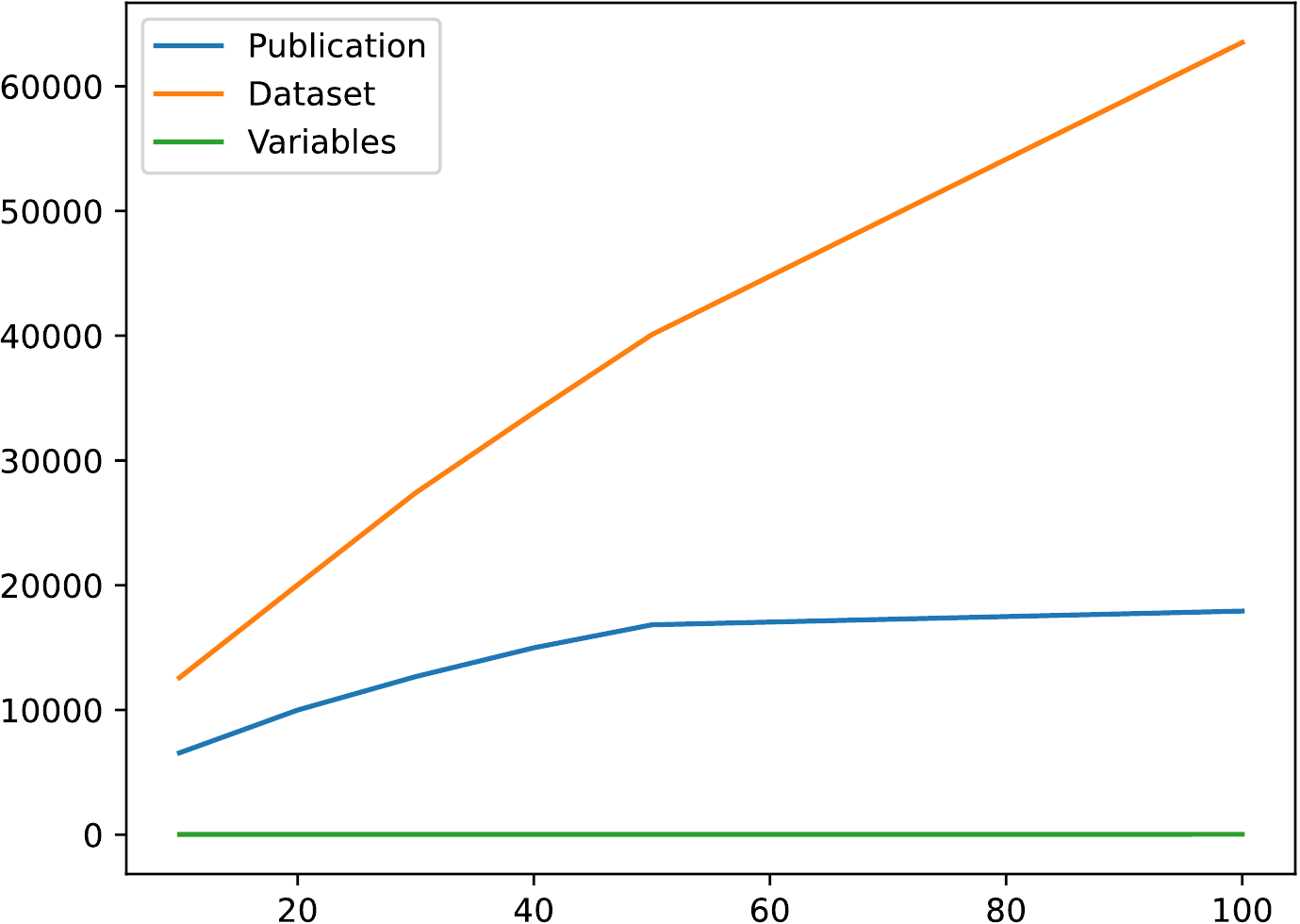}}
  \subfloat[Changes in standard deviation of r(d)] {\label{subfig:std}\includegraphics[width=0.253\textwidth]{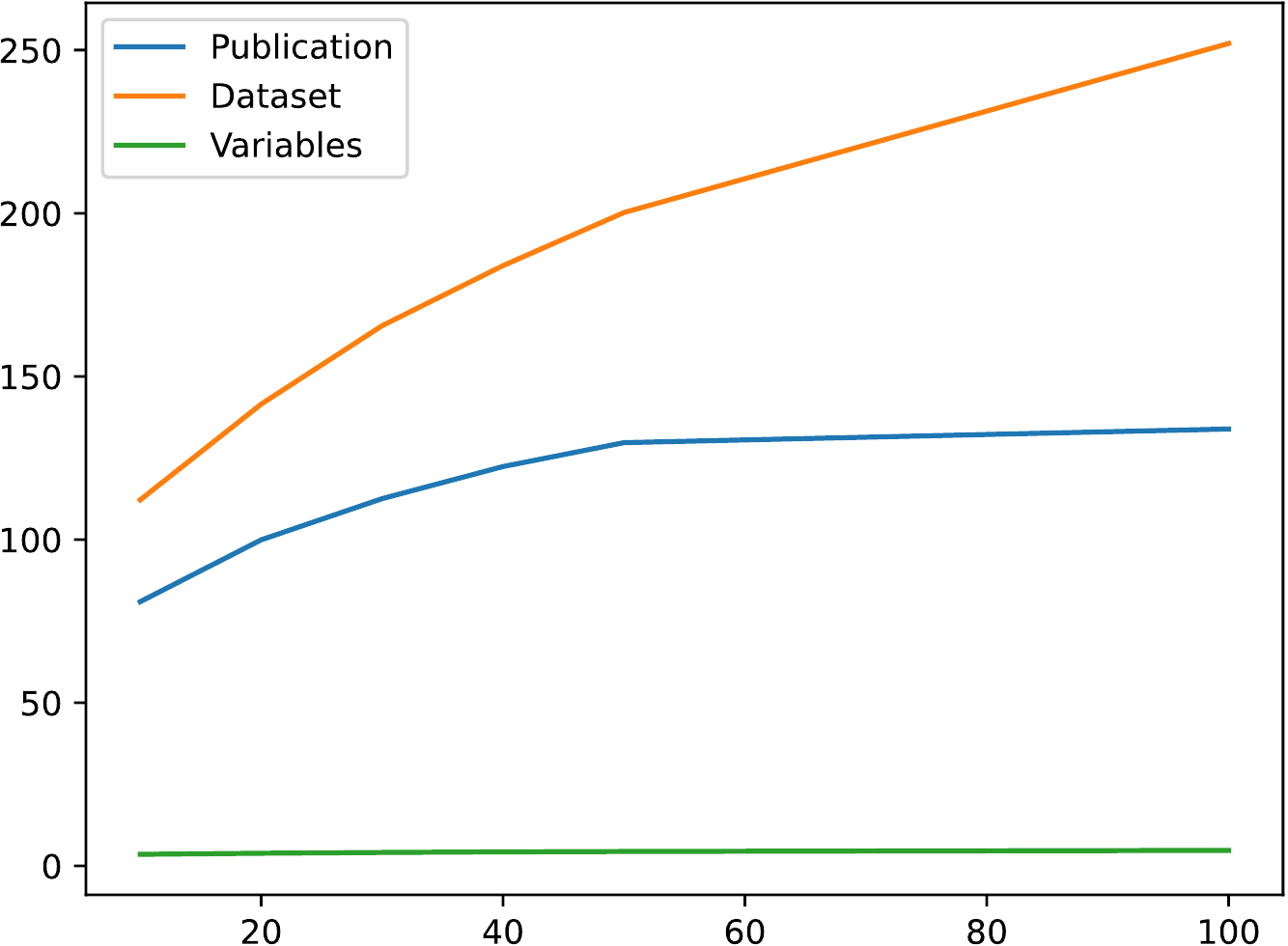}}\\
  \caption{Graphical representation of the change in various statistical measures of the observed distribution of retrievability scores. The mean, geometric-mean, variance and standard deviation of the distribution of retrievability scores of publication (in blue), dataset (in orange), and variables (in green) are presented.
\label{fig:rat-values-stats}}
\end{figure*}

\subsection{Experimentation}\label{subsec:exp}
As explained in Section~\ref{sec:rel-work}, the retrievability of a document is a measurement of how likely the document will be retrieved by \emph{any} query submitted to the system\footnote{By a system, we are referring to the organization of the collection, along with a retrieval model to be used for retrieval for a given query.}.
Hence, the study of retrievability in a collection of documents requires rigorous retrieval with a set of diversified queries to cover all topics discussed in the collection.
In other words, the retrievability of the documents should be calculated considering all sorts of queries submitted to the system.
However, an infinite number of queries are possible to be answered by a collection of free-text queries.
To cover all the topics, a traditional approximation is to simulate a set of queries randomly, accepting the risk of erratic queries not aligned with the real scenario~\citep{ret-cikm08,TraubSOHVH16}.
With the availability of a query log, the process of query generation can be made more formalized and streamlined to consider the actual queries submitted by real users.
For the study reported in this article, we utilize the query log presented in Section~\ref{subsec:datasets}.

As reported in the earlier study, the retrievability distribution in a collection depends on the employed retrieval model~\citep{ret-cikm08}.
Following the findings by Azzopardi and Vinay, we use BM25 as the retrieval model~\citep{bm25}.
Particularly, we use the implementation available in Elasticsearch\footnote{\url{https://www.elastic.co/}} which uses Lucene\footnote{\url{https://www.lucene.apache.org/}} as the background retrieval model.
Following Equation~\ref{eq:ret1}, the retrievability of a document depends on the selection of the rank cutoff value ($c$) - a rank threshold to indicate how deep in the ranked list are we going to explore before finding that document.
Considering the model employed for retrieval and the set of all queries $\mathsf{Q}$ as fixed, $c$ is the only parameter in calculating the retrievability.
For a query $q$, setting a lower value to $c$ will reduce the number of documents being considered retrievable because $f(k(d,q),c)$ will be $1$ only if $k(d,q) \leq c$ (see Equation~\ref{eq:fk}).
Having a higher value of $c$ will allow more documents to be considered retrievable reducing the overall inequality.
In this study, we have varied the value of $c$ in the range $10$ to $100$ in steps of $10$ and have analyzed the observations which are reported in the next section~\footnote{\ijdl{All codes are available here: \url{http://u.pc.cd/vzKctalK}}}.

\begin{figure*}[t]
  \centering
  \subfloat[Publication] {\label{fig:lc_pub_ret}\includegraphics[scale=0.37]{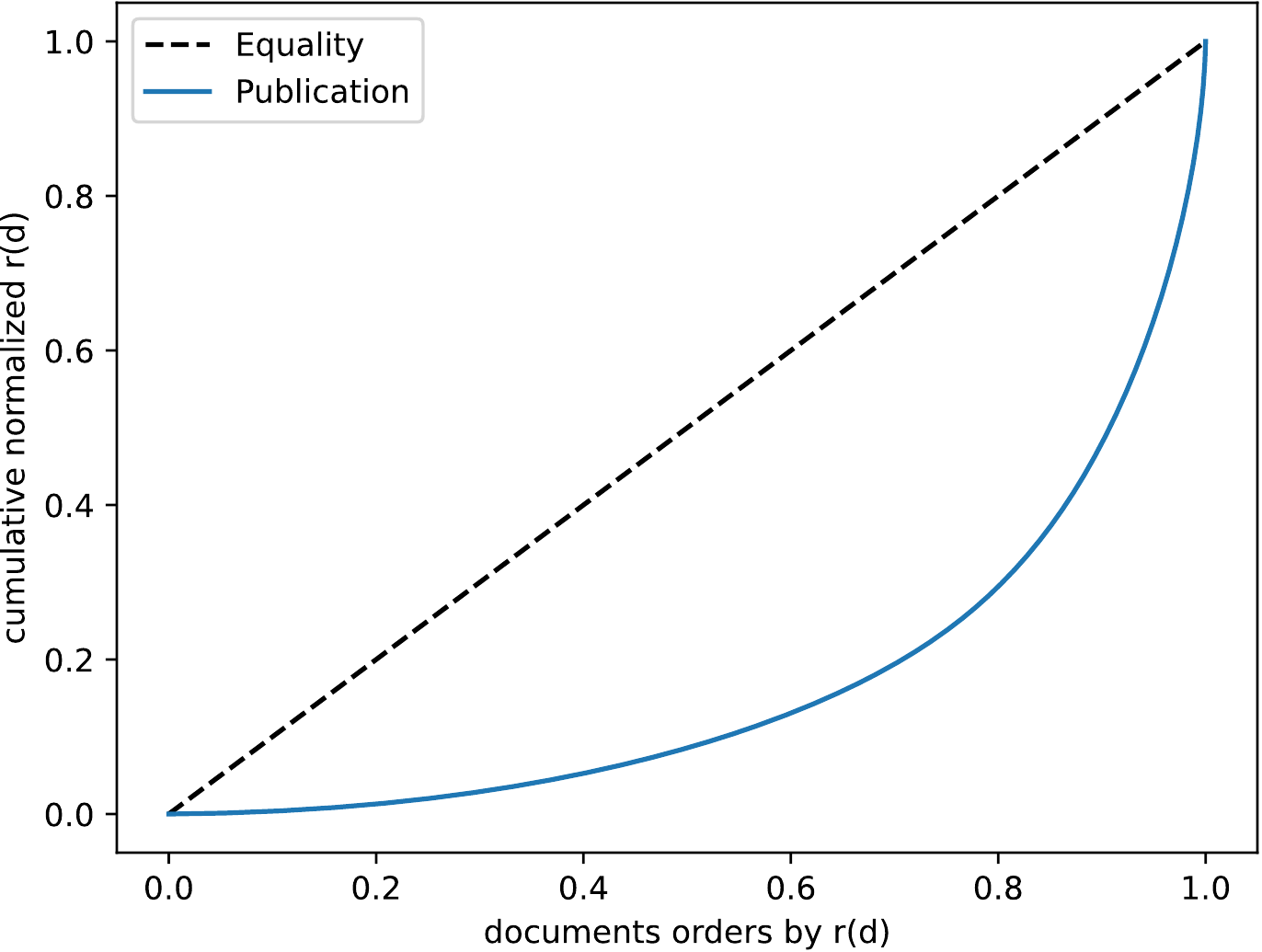}}
  \subfloat[Dataset] {\label{fig:lc_rd_ret}
\includegraphics[scale=0.37]{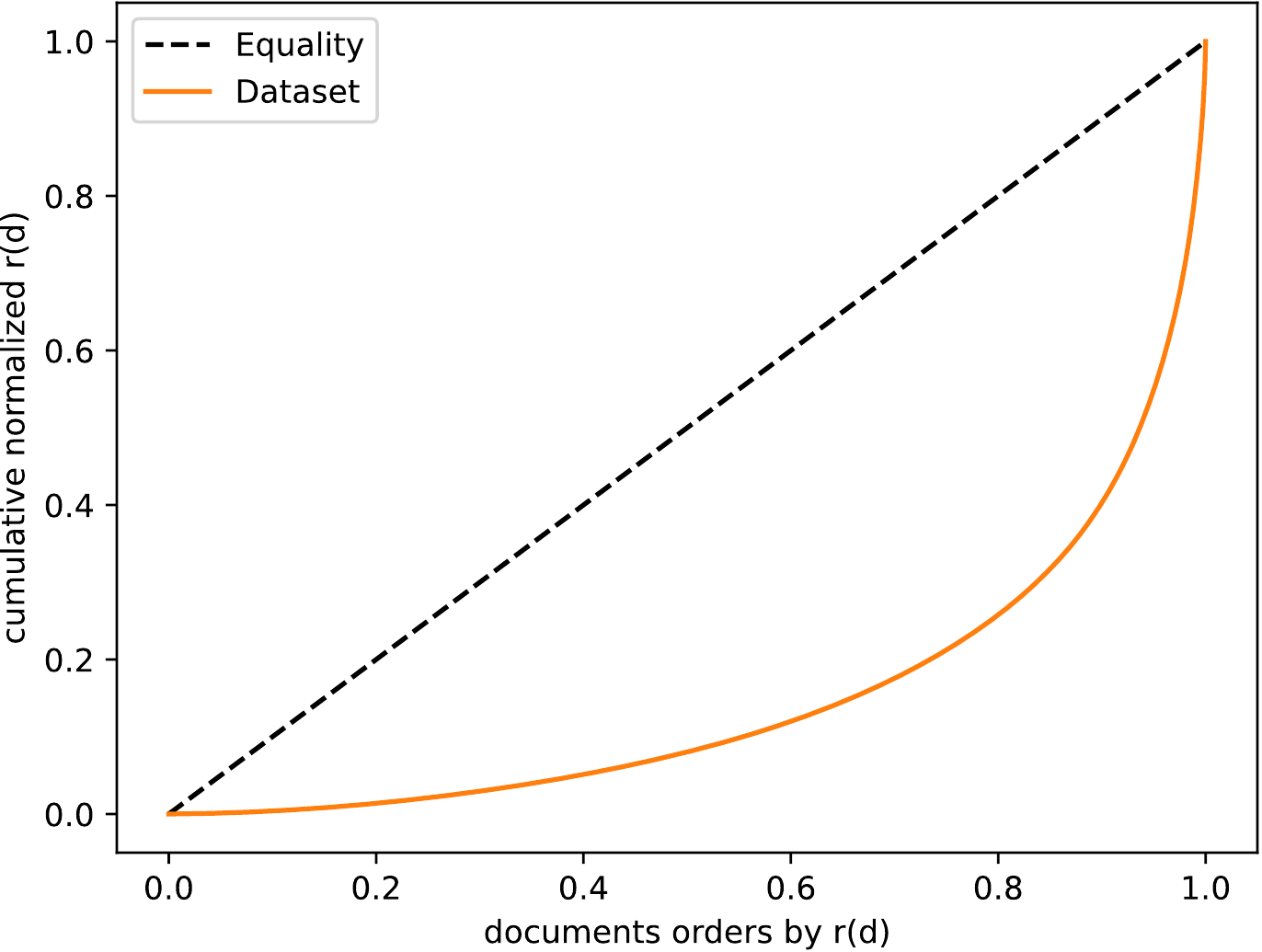}}
  \subfloat[Variables] {\label{fig:lc_var_ret}
\includegraphics[scale=0.37]{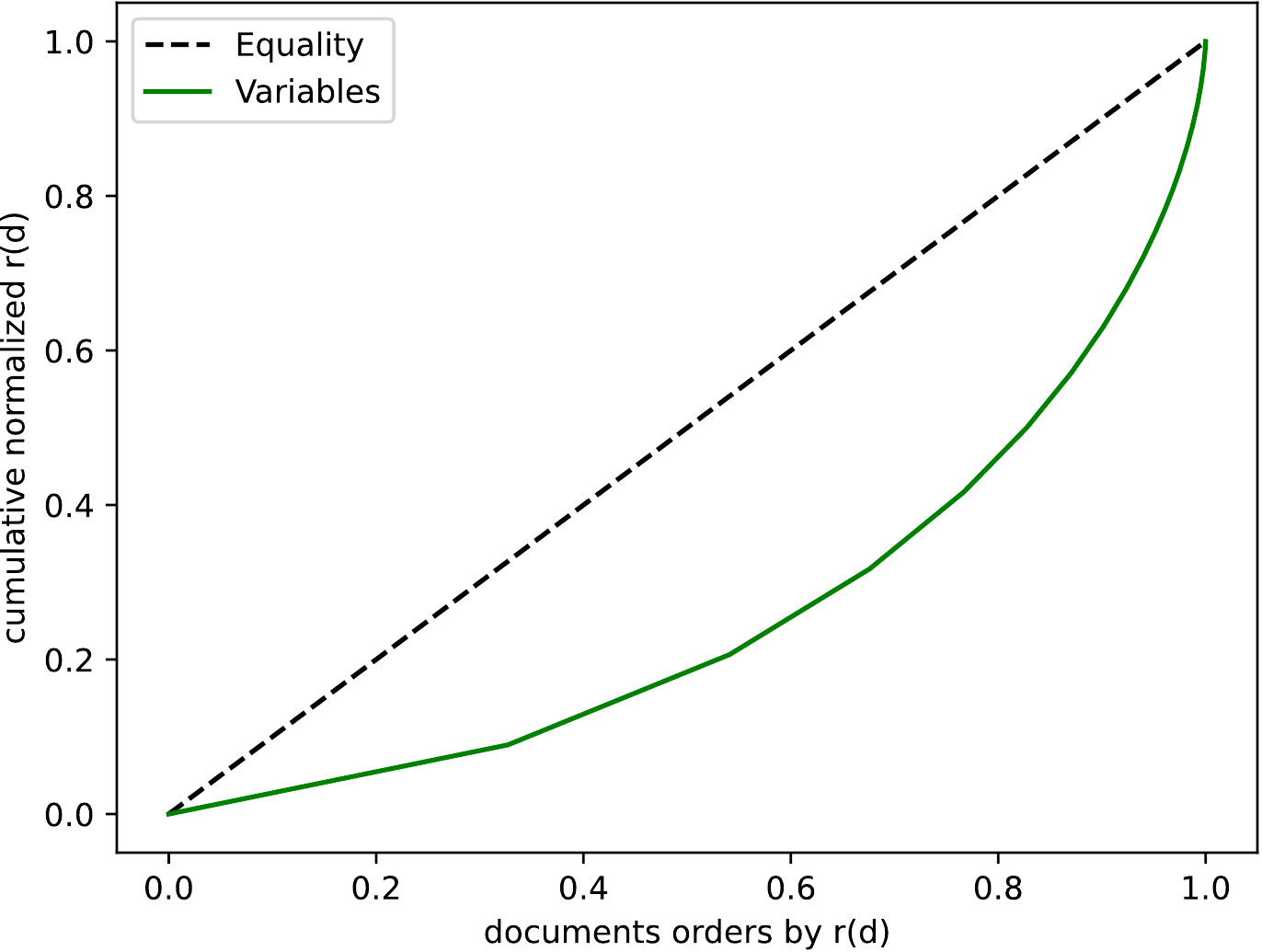}}
    \caption{The Lorenz curve with the retrievability (rank cutoff set to 100). The straight line going through the origin (in black) indicates the equality, that is, when all the documents are equally retrievable.
    }
    \label{fig:ret}
\end{figure*}

\subsection{Observation and analysis} \label{subsec:ana_ret}

\ijdl{
We start this section with describing different statistical properties of the retrievability distribution of items (from all the three different document types that we experimented with) when the value of $c$ is varied.
The mean ($\mu$), geometric mean (g-$\mu$), variance ($\sigma^2$), and standard deviation ($\sigma$) of the retrievability score distributions on different types (publication, dataset and variable) are given in Table~\ref{tab:ret-values-stats}.
In general, it can be noticed that all the statistical measures for datasets are far more diverse than the other categories.
On varying the value of $c$ from $10$ to $100$, we observe a change of more than 140\% and 220\% in mean retrievability scores in case of publication and dataset respectively while only 45\% change is noticed in case of variables.
In comparison to our earlier work~\citep{royjcdl22}, we can see these changes in the retrievability scores are moderate and are not as substantial as seen before.
Note that we have excluded repeated queries from the interaction log in this work which were considered in~\cite{royjcdl22}. 
This indicates that there is a significant number of repeated queries submitted into the system that had contributed to the momentous change reported earlier resulting in a vast diversity in retrievability scores (see~\cite{royjcdl22}, Table 2).
Similar trends are recorded for variance and standard deviation as well when computed using the distribution of $r(d)$ on all three categories with different $c$ values.
From Table~\ref{tab:ret-values-stats}, we can conclude that most of the statistical measurements (specifically mean, variance, and standard deviation) are higher for the datasets than publications.
In comparison, the geometric mean (\textbf{g-$\mu$} in Table~\ref{tab:ret-values-stats}) is seen to be higher for publications than datasets at the lower rank cut-offs. 
However, the geometric mean of retrievability of datasets surpasses that of publications at the rank cut-off 100.
Combining the observation that can be drawn from geometric-mean values together with the other statistics, we can perceive that for some dataset items, the retrievability values are extensive (popular datasets retrievable by a number of queries); at the same time, there are datasets with poor $r(d)$ values that are rarely retrieved through the submitted queries.
The first category of datasets are contributing to the high mean of $r(d)$, which is consistent across different $c$ values, while the datasets of the second category cause the geometric-mean to fall.
For the variables, we report all these measures are noticeably smaller than for publications and dataset.
The reason behind this is the relatively small number of queries of the variable category compared to the other types; 
as a result, the variables in general are selected for less number of queries in comparison to other categories.
These variations are presented graphically in Figure~\ref{fig:rat-values-stats}.
}
%

\ijdl{
As proposed in~\cite{ret-cikm08} and used in our earlier work~\citep{royjcdl22}, we utilize the Gini coefficient ($G$) to quantify the variation in retrievability scores, and Lorenz curve to graphically represent the disparity in retrievability among the items in different categories.
Figure~\ref{fig:ret} plots the Lorenz curve  with the $r(d)$ scores computed separately for publications, datasets and variables.
To consider the highest coverage, we set the rank cut-off $c$ to $100$ while plotting the $r(d)$ values\footnote{Similar trends are observed with $c$ set to lower values.}. 
From the Figure~\ref{fig:ret}, it is seen that retrievability of datasets (presented in Figure~\ref{fig:lc_rd_ret}) is more imbalanced than the other two types with Gini coefficient $0.7000$.
Also, variables are seen to be the closest to the equality (in Figure~\ref{fig:lc_rd_ret}) attaining a Gini coefficient of $0.4806$.
}

\begin{figure}
    \centering
    \includegraphics[scale=0.5]{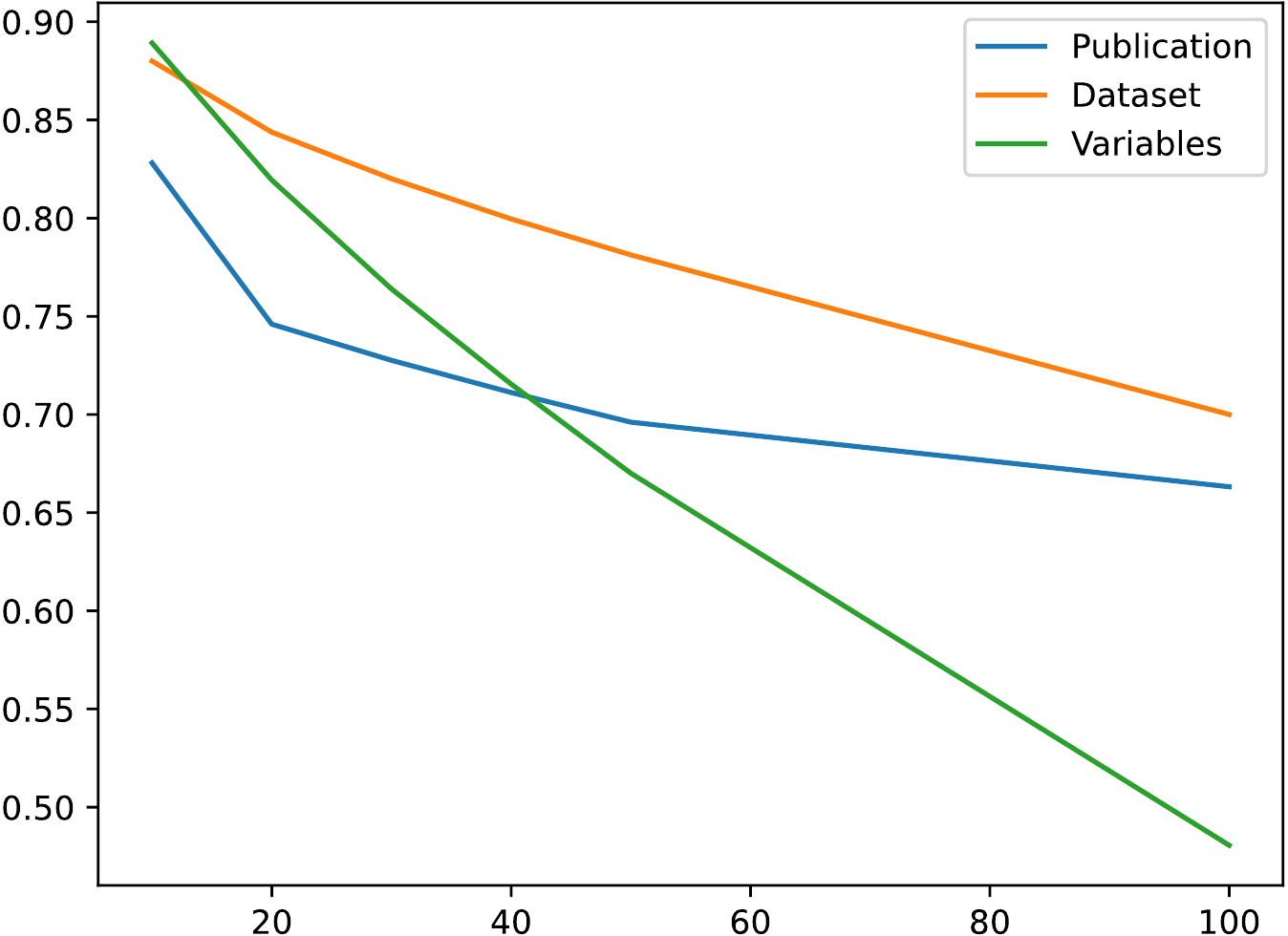}
    \caption{The change in Gini coefficient when the rank cut-off is varied in the range from 10 to 100. The blue line indicates the publication while dataset is specified by the orange curve.}
    \label{fig:change-g_with-c}
\end{figure}

\ijdl{
As discussed in Section~\ref{sec:rel-work}, the retrievability score of documents escalates with higher values of $c$; consequently, the overall retrievability-balance of the collection also changes positively bringing in the curve close to the equality.
To empirically see this variation, Gini coefficients attained at different rank cut-offs are presented in Table~\ref{tab:gini-change} which is also graphically displayed in Figure~\ref{fig:change-g_with-c}.
From the table, it can be noticed that the fall in $G$ for variables (green curve in Figure~\ref{fig:change-g_with-c}) is more than 45\%.
From a severe unequal distribution with $G$ having $0.8281$ till rank $10$ (highest among all the categories), the Gini value falls sharply at $0.4806$ when the rank cut-off is set to 100.
This indicates that more variables are discernible if the ranked list is explored beyond the top position.
}

\ijdl{
Additionally, we report the percentage of total items retrieved while changing $c$ in Table~\ref{tab:gini-change}.
Note that more than 92\% of publication are retrieved within the top $10$ positions while only 58\% and 10\% items respectively from the category dataset and variables are retrieved within the same rank cut-off.
Increasing the value of $c$, it is noticed that more than $98\%$ documents are retrievable within the top $100$ ranked documents by all the queries for both publication and dataset.
The significant change in the percentage of retrieved documents of type dataset indicates that searching for datasets is more complex than publications; a deeper ranked list traversal might be essential to find a relevant dataset.
Note that only half of the items from variables category (specifically 50.43\%) are retrieved within the top $100$ positions although the Gini value indicates more balance in retrievability ($G = 0.4806$).
This leads to an interesting observation: as reported in Table \ref{tab:ret-values-stats}, the average retrievability scores for variables are significantly smaller ($r(d)=3.67$ at cut-off $100$), the difference in not being retrieved (having $r(d)$ = 0) and retrieved with average retrievability score  is merely a small value.
Due to this seemingly-inconsequential difference in $r(d)$ score, the Gini is not affected significantly.
However, these variables, which are not retrieved at all, lowers the percentage of retrieved items.
}

\begin{table*}[t]
\resizebox{0.99\textwidth}{!}{
\begin{tabular}{c ccc l cccccc}
\hline
\textbf{Rank} & \multicolumn{3}{c}{\textbf{Gini coefficient}} & &  \multicolumn{6}{c}{\textbf{Retrieved}} \\
\cmidrule{2-4} \cmidrule{6-11} 
\multicolumn{1}{l}{\textbf{cutoff}} & \multicolumn{1}{l}{\textbf{Publication}} & \multicolumn{1}{l}{\textbf{Dataset}} & \multicolumn{1}{l}{\textbf{Variable}} & & \multicolumn{1}{l}{\textbf{Publication}} & \multicolumn{1}{l}{\textbf{\%}} & \multicolumn{1}{l}{\textbf{Dataset}} & \multicolumn{1}{l}{\textbf{\%}} & \multicolumn{1}{l}{\textbf{Variable}} & \multicolumn{1}{l}{\textbf{\%}} \\ \hline
10                                       & 0.8281                                   & 0.8800                                       & 0.8892                           &     & 110666                                   & 92.15                           & 37554                                & 58.31                           & 53799                                 & 10.28                           \\
20                                       & 0.7460         & 0.8438                                     & 0.8194                            & & 116322                                   & 96.86                           & 46437                                & 72.10                            & 89959                                 & 17.20                            \\
30                                       & 0.7276                                   & 0.8201                                     & 0.7640                              &   & 118050                                   & 98.30                    & 51160                                & 79.44                           & 118961                                & 22.74                           \\
40                                       & 0.7112                                   & 0.7996                                     & 0.7155                             && 118819                                   & 98.94                           & 54503                                & 84.63                           & 144393                                & 27.60                            \\
50                                       & 0.6961                                   & 0.7813                                     & 0.6701                               & & 119259                                   & 99.31                           & 56761                                & 88.13                           & 167691                                & 32.06                           \\
100                                      & 0.6632                                   & 0.7000                                        & 0.4806                                && 119847                                   & 99.80                    & 63735                                & 98.96                           & 263801                                & 50.43                      \\ \hline    
\end{tabular}
}
\caption{Change in Gini coefficient when the rank cut off is increased. Also the number and percentage of documents retrieved of type Publication Dataset and Variable are presented.}\label{tab:gini-change}
\end{table*}

\subsection{\ijdl{Comparing influence of query popularity bias}}

Considering a real-life query log, there is an obvious possibility of having more than one entry for the popular queries.
While computing the retrievability, the items retrieved by those repeated queries get a boost in the retrievability score due to the popularity bias of the queries.
To understand the influence of this query popularity bias, in this section, we report relationship between the retrievability scores of the items computed with $i)$ $Q_r$ - the interaction log containing \emph{repeated} queries, and $ii)$ $Q_u$ - the query log with only the \emph{unique} queries\footnote{Note that as the system may evolve with new documents being added into the index, the exact ranked list produced for the same query submitted at two different times may differ. However, we have ignored the evolving nature of the index and have considered the latest snapshot of the index to perform the retrieval.}.
Particularly, we report how disjoint the documents with the highest retrievability scores are when the retrievabilities are computed with the two types of queries separately.
If the documents are ordered by their retrievability scores, we get two individual ranked lists of documents each when $Q_r$ and $Q_u$ are employed.
In order to compare and contrast the lists produced by the two types of query lists, we adapt three ways to quantify the difference:\\
\begin{table}[h]
    \centering
\begin{tabular}{ ccc c} 
 \hline
 \textbf{Top items} & \multicolumn{3}{c}{\textbf{Jaccard's coefficient}} \\ 
 \cmidrule{2-4}
 \textbf{considered} & \textbf{Publication}& \textbf{Dataset} & \textbf{Variable} \\ 
 \hline
 \multirow{2}{*}{\textbf{1000}} & \multirow{2}{*}{0.1025} & \multirow{2}{*}{0.1287} & \multirow{2}{*}{0.3199}\\ \\
 \multirow{2}{*}{\textbf{5000}}& \multirow{2}{*}{0.2917} & \multirow{2}{*}{0.2606} & \multirow{2}{*}{0.4319} \\ \\ 
 \multirow{2}{*}{\textbf{10000}}& \multirow{2}{*}{0.3896} & \multirow{2}{*}{0.3546} & \multirow{2}{*}{0.5473} \\ \\ 
 \multirow{2}{*}{\textbf{20000}}& \multirow{2}{*}{0.4584} & \multirow{2}{*}{0.4821} & \multirow{2}{*}{0.6353} \\ \\ 
 \multirow{2}{*}{\textbf{50000}}& \multirow{2}{*}{0.5756} & \multirow{2}{*}{0.8383} & \multirow{2}{*}{0.8897} \\ \\ 
 \hline
\end{tabular}
    \caption{The Jaccard's coefficient (set-based similarity) between the ranked lists of items obtained with different query sets $Q_r$ and $Q_u$ are reported. The first column indicates the number of top retrievable items considered to compute the similarity.}
    \label{tab:set-based}
\end{table}

\begin{itemize}
    \item \textbf{Set-based:} 
    We compute the Jaccard's coefficient between the two lists ranked by their retrievability scores till different rank cut-offs.
    Particularly, the first 1K, 5K, 10K, 20K and 50K top-ranked items are considered and their set-based overlap is computed.
    The results are reported in Table~\ref{tab:set-based}.
    From the results, we can see that overlap in items having the highest 1K retrievability scores are 10\% and 12\% respectively for the categories publication and dataset.
    However, around 31\% overlap is observed for the variable category among top 1K items.
    The Jaccard's coefficient changes swiftly for all the categories when higher number of items are considered.
    This indicates that the diversity between the two types of ranked lists are significant for all the three categories of items.
    
    \item \textbf{Correlation-based:} 
    Further, we compare the two ranked lists in terms of their correlations.
    Based on the discordant and concordant pairs, we compute the Kendall's $\tau$ correlation coefficient.
    Additionally, the Spearman's rank correlation is also assessed and reported in  Table~\ref{tab:correl} for all three categories.
    Considering these measures, we note that the rank correlations indicate an imperceptible relation between the two lists for all of the types while the most diverse results are observed in the case of publication category.
    For variables, the correlations are noted to be higher as compared to the other types whereas it is too inconsiderable for the other types. 

    \item \textbf{Rank overlap-based:} 
    The correlation-based measures suffer from certain limitations such as the lists needing to be conjoint and the measurement does not consider the position where the disagreements are happening; that is, the measure does not discriminate between mismatch at top position and at later positions.
    As an alternative,~\cite{rbo} proposed a ranked-biased overlap measure (RBO) that weights the difference considering the position at which they are occurring.
    Mathematically, the RBO between two ranked lists $S$ and $T$ is computed as:
    \begin{equation}
    RBO(S, T, p) = (1-p) \sum_{d=1}^{\infty}p^{d-1}\cdot A_d
    \end{equation}
    In the Equation, $d$ is the depth of the list, $p$ is a weighting factor (between 0 and 1) and $A_d$ is the common items at depth $d$ divided by the depth $d$ itself.
    Following~\citeauthor{rbo}, we have set the weight parameter $p$ to $0.9$.
    The RBO-based similarity between the two types of results is reported in Table~\ref{tab:correl}. 
    Again, it is prominent from the results that the dissimilarities between the rank of the items based on their retrievability scores are noteworthy, particularly for the publication and dataset categories.
\end{itemize}

From the dissimilarities between the two ranked items of all three categories, it can be concluded that the popularity bias of queries affects the retrievability irrespective of the type.
Out of the three categories, comparatively the least influence by this bias is observed for items belonging to the variable categories.
The retrievability of items from the publication and dataset categories are noted to be the most impacted with less than 13\% common items being observed among the top 1K.

\begin{table}[h]
    \centering
\begin{tabular}{ ccc c} 
 \hline
 & \multicolumn{3}{c}{\textbf{Correlation coefficient}} \\ 
 \cmidrule{2-4}
 \textbf{Measure} & \textbf{Publication}& \textbf{Dataset} & \textbf{Variable} \\ 
 \hline
 
 \multirow{2}{*}{\textbf{Kendall's $\tau$}}    & \multirow{2}{*}{0.0279} & \multirow{2}{*}{0.0789} & \multirow{2}{*}{0.1275}  \\ \\
 \multirow{2}{*}{\textbf{Spearman's $r$}}& \multirow{2}{*}{0.0390} & \multirow{2}{*}{0.1179} & \multirow{2}{*}{0.2267}  \\ \\ 
 \multirow{2}{*}{\textbf{RBO}}& \multirow{2}{*}{0.4594} & \multirow{2}{*}{0.6211} & \multirow{2}{*}{0.7119}  \\ \\ 
 \hline
\end{tabular}
    \caption{The rank correlation based (Kendall's $\tau$ and Spearman's $r$) and rank-bias based (RBO) similarities between the ranked lists of items obtained with different query sets $Q_r$ and $Q_u$ are reported.}
    \label{tab:correl}
\end{table}

\section{From Retrievability to Usefulness}\label{sec:usefulness}
Usefulness was introduced in \cite{cole2009usefulness} and designed initially as a criterion for the evaluation of interactive search systems.
The \emph{usefulness} of a document can be defined as how often the document is retrieved and \emph{exported} (see Section \ref{subsec:datasets})  by the end user. 
Of course the concept of usefulness can only reliably be recognized by relevance judgements submitted by the user for a given query and the relevance of a document may also depend on the perspective of the user which may vary across users and different points in time.
Without an explicit relevance judgement, the approximation of usefulness of documents can not be reliably accomplished. 
Considering the availability of the export and utilisation information from the query log, we can define the usefulness of a document ($u(d)$) by the following equation:

\begin{equation}\label{eq:usefulness}
    u(d) = \sum_{q\in \mathsf{Q}}w_q \cdot g(d,q)
\end{equation}

In Equation~\ref{eq:usefulness}, the weight of the query ($w_q$) can be defined in a similar way as defined in retrievability (Equation~\ref{eq:ret1}).
The usefulness of a document may also depend on the \textit{difficulty} of the query~\citep{query-dif2006carmel,dif-book2010carmel}\footnote{A query can be considered as \emph{difficult} if the top ranked documents are mostly non-relevant in which scenario, the user has to go deep down the ranked list to get the document addressing the query~\cite{dif-book2010carmel}.}.
A document $d$ should be considered more useful if it is retrieved and consumed following a query $Q$ than any other document, say $d'$ with an associated query $Q'$ which is relatively easier than $Q$ (i.e. $difficulty(Q)>difficulty(Q')$).
Hence, we extend the definition of the weight of the query taking into account a difficulty factor in Equation~\ref{eq:w_q_d}.

\begin{equation}\label{eq:w_q_d}
    w'_q = w_q * h(q)
\end{equation}
where the function $h(q)$ represents the difficulty of the query $q$. 
The function $g(\cdot)$ in Equation~\ref{eq:usefulness} indicates usefulness in terms of relevance of the document $d$ for the query $q$.
Mathematically, $g(\cdot)$ can be defined as follows:
\begin{equation}\label{eq:gk}
  g(d,q) = rel(d,q)
\end{equation}

The function $rel(d,q)$ in Equation~\ref{eq:gk} indicates the relevance of $d$ for the query $q$.
It works, in the same way, $f(k(d,q),c)$ is defined in Equation~\ref{eq:fk} considering a binary relevance (that is $d$ can be either relevant - $rel(d,q)=1$, or non-relevant - $rel(d,q)=0$ to the query $q$).

\ijdl{Informally speaking, the usefulness of a document can be generally stated as the number of queries for which, it is exported (i.e. consumed) by the user. 
Considering a SERP without any duplicate documents, the usefulness can be further simplified as the count of exportation of the document.}


\begin{figure*}[t]
  \centering
  \subfloat[Publication] {\label{fig:lc_pub_use}
  \includegraphics[scale=0.37]{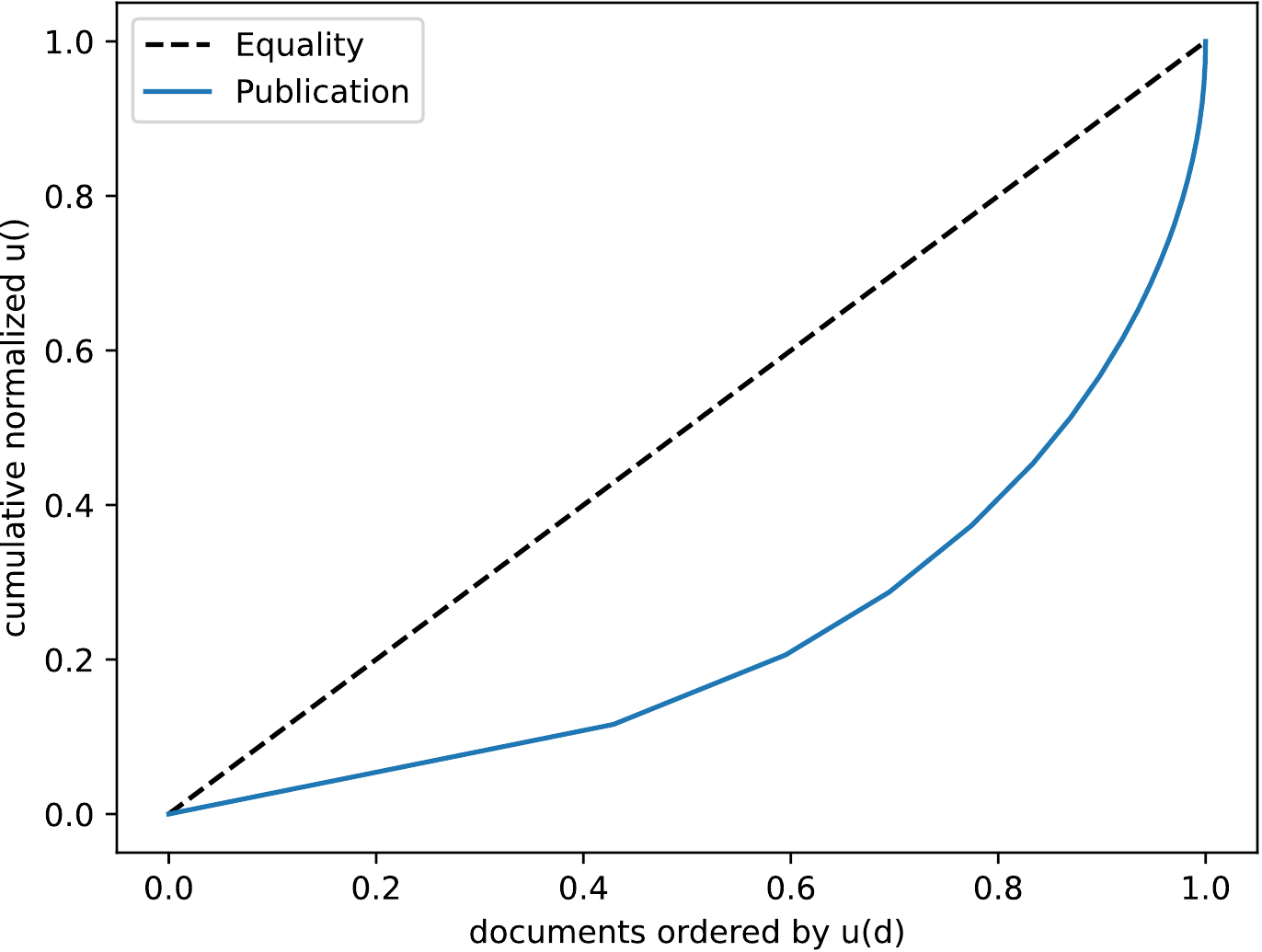}}
  \subfloat[Dataset] {\label{fig:lc_rd_use}
    \includegraphics[scale=0.37]{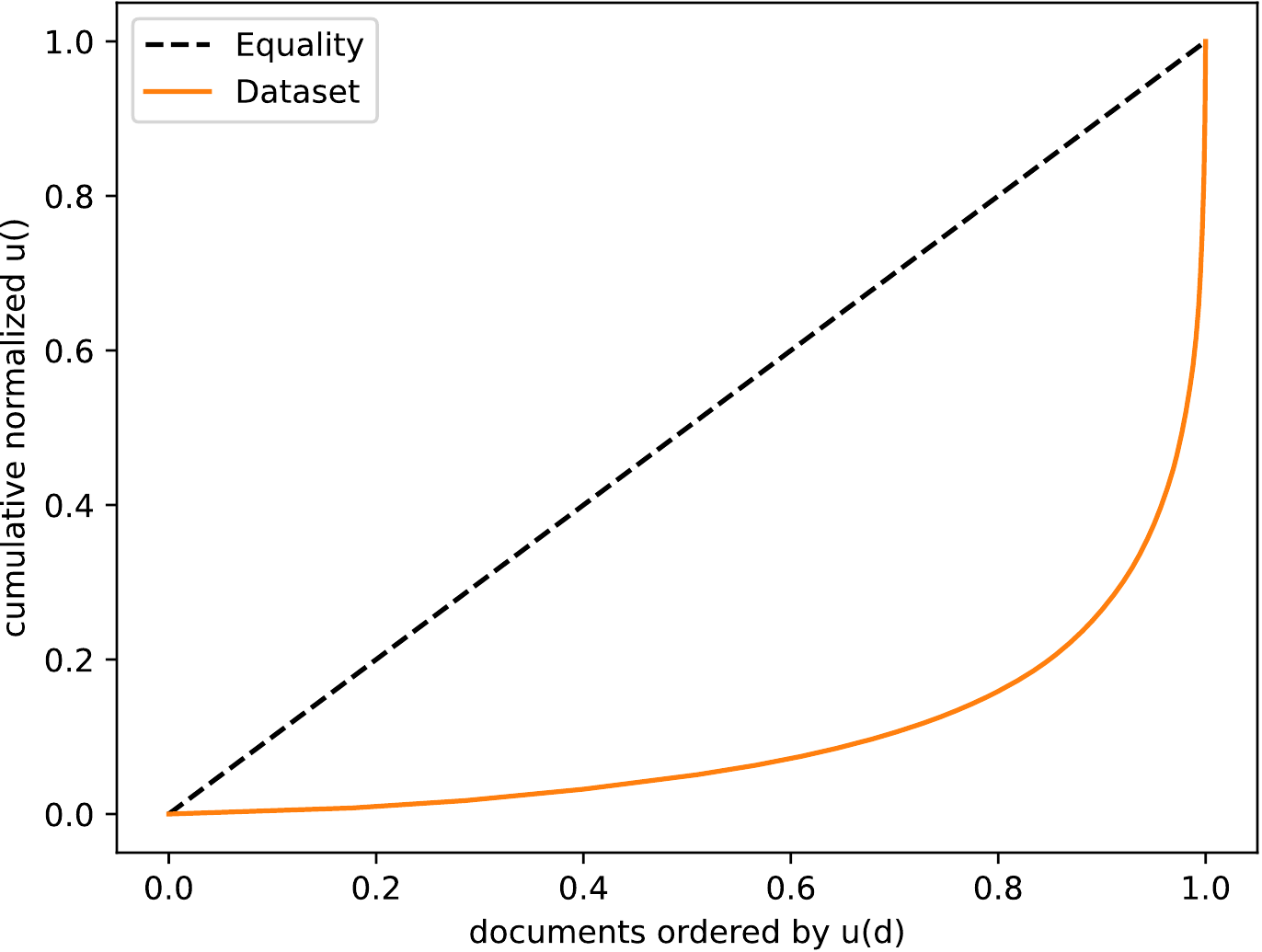}}
  \subfloat[Variables] {\label{fig:lc_vr_use}
    \includegraphics[scale=0.37]{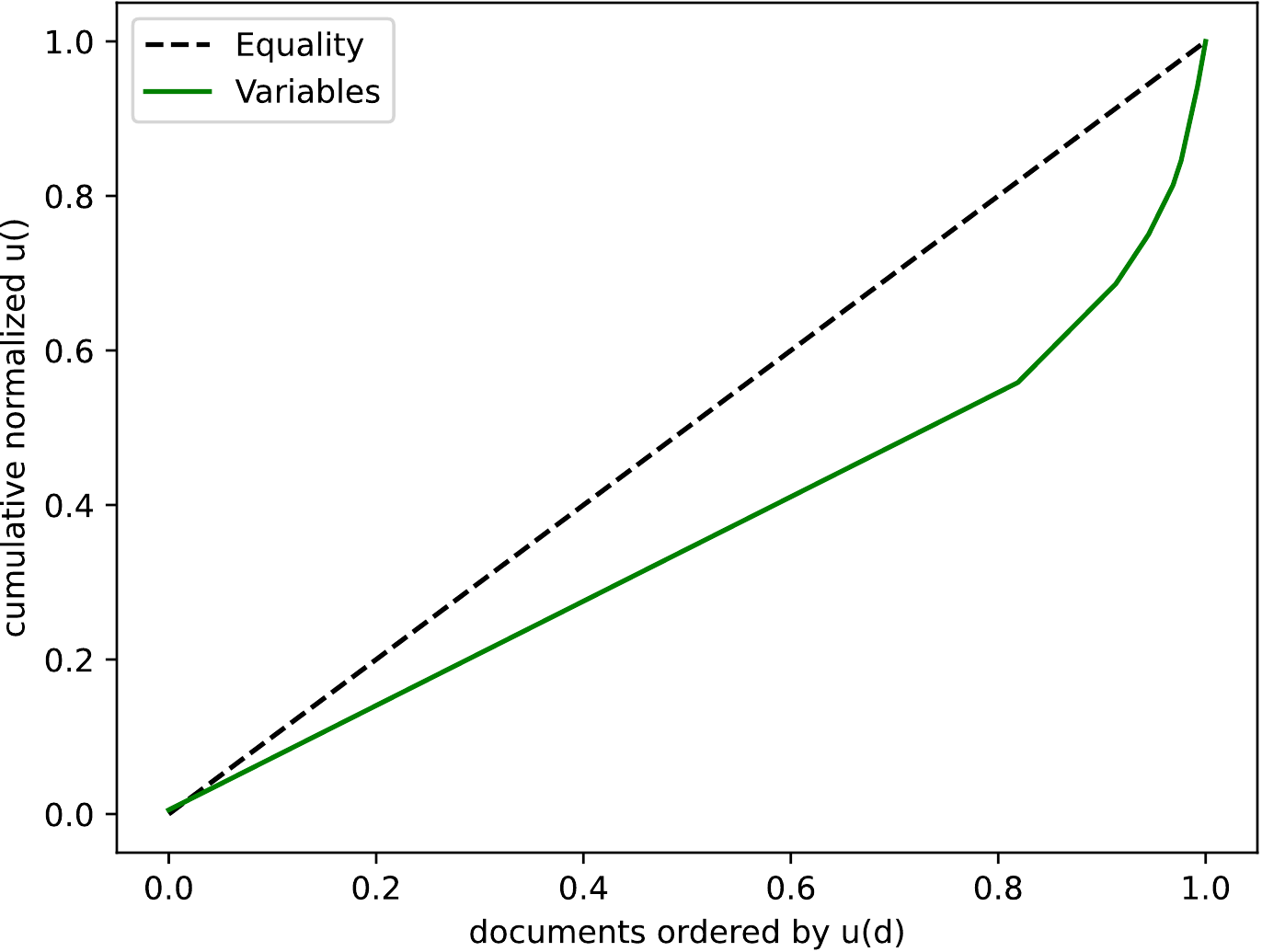}}
    \caption{Plotting Lorenz curves for usefulness values. The straight line going through the origin (in black) indicate the \emph{equality}, that is, when all the documents are equally useful.
    The blue (Figure~\ref{fig:lc_pub_use}) and orange (Figure~\ref{fig:lc_rd_use}) curves respectively specify the publication and dataset, while variable is indicated by the green curve (Figure~\ref{fig:lc_vr_use}).
    }
    \label{fig:use}
\end{figure*}

\subsection{Experimentation} \label{subsec:exp_usefulness}
As presented and argued earlier in Section~\ref{sec:usefulness}, the signal of document consumption by the user is essential in order to compute the usefulness of documents.
We utilize the information stored in the interaction log of the integrated search system \textit{GESIS Search} as the indication of document consumption by the user.
Particularly, the usefulness is determined on the basis of implicit relevance feedback from the \emph{export} interactions (see Section~\ref{subsec:datasets}).
The difficulty of the query is kept as constant ($h(q)$ in Equation~\ref{eq:w_q_d} set to $1$) in this study and further study in this regard has been left as part of future work.

\subsection{Observation and analysis} \label{subsec:ana_usefulness}
The experimental results on usefulness are graphically presented in Figure~\ref{fig:use} where a pair of Lorenz curves are displayed with the usefulness of the documents of type publication, dataset and variable. 
From Figure~\ref{fig:lc_vr_use}, we can observe that the usefulness distribution of variables is close to being equally distributed as compared to the other types.
In comparison, the similar distribution of datasets (presented in Figure~\ref{fig:lc_rd_use}) is observed to be more skewed with an evident inclination towards certain items being more useful.
The corresponding Gini coefficient of the distributions is presented in Table~\ref{tab:gini_usefulness} where the value of $G$ for the usefulness of dataset distribution is seen to be almost three times greater than the variables.
The difference in publications and datasets is also evinent.
This observation clearly highlights that a few datasets are more useful than the rest, whereas the usefulness distribution of the variables is considerably close to being uniform.
In the case of publications, the distributions are also observed to be similar to that of variables which are close to uniformity.

\begin{table}[h]
    \centering
\begin{tabular}{ ccc c} 
 \hline
 & \textbf{Publication}& \textbf{Dataset} & \textbf{Variables}\\ 
 \hline
 \textbf{Gini}  & \multirow{2}{*}{0.3160} & \multirow{2}{*}{0.8031} & \multirow{2}{*}{0.2876} \\
 \textbf{coefficient} &&&
 \\ \hline
\end{tabular}
    \caption{The Gini coefficient computed with the distribution of usefulness of the publication, dataset and variables. A higher Gini coefficient (upper bound 1.0) indicates an uneven distribution of usefulness.}
    \label{tab:gini_usefulness}
\end{table}

\section{\ijdl{Conclusion and future work}}\label{sec:conclusion}

In~\cite{royjcdl22}, we have reported a significant difference in retrievability of items belonging to various categories in the integrated search system \textit{GESIS Search}.
We particularly focused on the types \textit{publications} and \textit{datasets} and concluded that there is a significant difference in the retrievability scores if the item belonged to the category of publication or dataset.
As an extension to that work, we have included another category to study the retrievability which is \textit{variables}.
Along with that, we have used a newer and larger version of interaction logs for our experimentation.
A noticeable difference in the experimental setup from our earlier work is that we have used a deduplicated version of the log.
That is, only the unique queries from the interaction log are considered excluding any repeated entries.
This ensures bypassing any query popularity bias, which may influence the retrievability of the items.

In this extended study, we observe similar phenomena on the newer data as well as on the variable type.
In response to \textbf{RQ1}, we have seen a significant popularity-bias with certain items being retrieved more often than others.
Particularly, it has been shown that certain items from the dataset category are more likely to be retrieved than the other items in the same category.
In contrast, the retrievability scores of items from variable or publication types are more evenly distributed.
For the \textbf{RQ2}, the intra-document selection bias is formalized using the common measures of Lorenz curve and Gini coefficient. 
In response to \textbf{RQ3}, we have observed that the distribution of document retrievability is more diverse for the datasets as compared to publications.
This can be attributed again to the popularity bias of certain items in the dataset category.
The earlier study used an interaction log not employing any deduplication of queries; as a result, the items retrieved for those popular queries (occurring frequently in the log) gain a boost in the computed retrievability scores.
In this paper, we have further included an explicit discussion and comparison of the retrievability scores of items in different categories when the query popularity bias is factored out by the deduplication of the queries.
In this connection, as a response to \textbf{RQ4}, we showed that there can be a positive influence of the query popularity bias on the distribution of the retrievability scores.

Further study on the measurement of usefulness (proposed in our earlier work~\citep{royjcdl22}) reveals a prominent diversity in the nature of consumption of items among the different types.
We notice that variables are close to having an equality in usefulness which is significantly disparate in publication and dataset categories.
Additionally, we have proposed a measurement of \emph{usefulness} of documents based on the signal of document consumption by the users after submitting a query to the system.
Experimenting with the variables, we observe that the usefulness of items in this category is closer to equality than items in the other categories.

The proposed usefulness metric indicates its popularity in terms of being consumed by the users. Hence one possible extension of this work will be to test the applicability of usefulness to improve retrieval performance. 
Incorporating the usefulness of documents as a feature in the learning to rank framework could actually boost the retrieval effectiveness.
In terms of presenting the results (SERP) to end users, usefulness can be used as a sorting measure to organise the retrieved items based on popularity.
Specifically, together with  the provision of presenting the  results sorted based on the recency or relevance, it can also be extended to provide an ordering based on how popularly the document is viewed by the users.

\paragraph{Acknowledgment}
This work was funded by DFG under grant MA 3964/10-1, the ``Establishing Contextual Dataset Retrieval - transferring concepts from document to dataset retrieval'' (ConDATA) project at GESIS. Dwaipayan Roy wants to acknowledge a research grant provided by the GESIS Research Gateway EUROLAB in summer 2022.

\paragraph{Conflict of interest}
Philipp Mayr is on the Editorial Board of the ``International Journal on Digital Libraries'' and guest co-editor of the special issue ``JCDL 2022''. In this case, the co-editors are handling the review process.

\bibliography{main}


\end{document}